\documentclass[useAMS,usenatbib,usegraphicx]{mn2e}
\usepackage{comment}

\newcommand{\br}{B$-$R }
\newcommand{\f}{[3.6] }
\newcommand{\rf}{R$-$\f }
\newcommand{\bc}{B$-$\f }

\title[Spitzer IRAC Photometry of GC Systems]{Extending the baseline:  Spitzer Mid-Infrared Photometry of Globular Cluster Systems in the Centaurus~A and Sombrero Galaxies}
\author[L. R. Spitler, D. A. Forbes and M. A. Beasley]{Lee R. Spitler$^{1}$\thanks{E-mail: lspitler@astro.swin.edu.au } Duncan A. Forbes$^{1}$ and Michael A. Beasley$^{2}$\\
$^{1}$Centre for Astrophysics \& Supercomputing, Swinburne University, Hawthorn, VIC 3122, Australia\\
$^{2}$Instituto de Astrof\'{i}sica de Canarias, Via Lactea, E-38200 La Laguna, Tenerife, Spain}

\begin{document}

\pagerange{\pageref{firstpage}--\pageref{lastpage}} \pubyear{2008}

\maketitle

\label{firstpage}

\begin{abstract}
Spitzer IRAC mid-infrared photometry is presented for the globular cluster (GC) systems of the NGC~5128 (``Centaurus~A'') and NGC~4594 (``Sombrero'') galaxies.  Existing optical photometric and spectroscopic are combined with this new data in a comprehensive optical to mid-IR colour catalogue of 260 GCs.    Empirical colour-metallicity relationships are derived for all optical to mid-IR colour combinations.

These colours prove to be very effective quantities to test the photometric predictions of simple stellar population (SSP) models.  In general, four SSP models show larger discrepancies between each other and the data at bluer wavelengths, especially at high metallicities.  Such differences become very important when attempting to use colour-colour model predictions to constrain the ages of stellar populations.  Furthermore, the age-substructure determined from colour-colour diagrams and 91 NGC~5128 GCs with spectroscopic ages from \citet{2008MNRAS.386.1443B} are inconsistent, suggesting any apparent GC system age-substructure implied by a colour-colour analysis must be verified independently.

Unlike blue wavebands, certain optical to mid-IR colours are insensitive to the flux from hot horizontal branch stars and thus provide an excellent metallicity proxy.  The NGC~5128 GC system shows strong bimodality in the optical R-band to mid-IR colour distributions, hence proving it is bimodal in metallicity.  In this new colour space, a colour-magnitude trend, a ``blue tilt'', is found in the NGC~5128 metal-poor GC data.  The NGC~5128 young GCs do not contribute to this trend.  In the NGC~4594 GC system, a population of abnormally massive GCs at {\it intermediate metallicities} show bluer optical to optical colours for their optical to mid-IR colours, suggesting they contain extended horizontal branches and/or are younger than typical GCs.  Analysis of optical to mid-IR colours for a ultra-compact dwarf galaxy suggests its metallicity is just below solar.
\end{abstract}


\section{Introduction}

Massive star cluster formation takes place during major star bursts.  The age substructure of a particular star cluster system therefore provides a historical record of significant star formation events in its host galaxy.  Globular clusters (GCs) are a class of massive star clusters with a mean mass of a few $10^{5} M_{\odot}$ and are notable for their compactness.  Observational evidences generally suggests the bulk of GC formation took place within roughly 2 gigayears of the Big Bang (see references in Brodie \& Strader 2006).   

Systems of GCs are known to host a blue and red GC subpopulations in terms of their photometric colour (e.g. Peng et al. 2006; Strader et al. 2006).  This has been interpreted as evidence for a bimodal metallicity distribution, hence most GC systems contain both a metal-poor (or blue) and metal-rich (red) GC subpopulation.  Given the close proximity of the GC formation epoch to the time when the universe was free of metals, the existence of two GC metallicity subpopulations likely implies that two distinct phases of GC formation took place at different epochs.  Obtaining absolute age dates of the two star formation events (and subsequent ones) can help constrain the detailed formation and assembly history of a galaxy.  This has encouraged efforts to age-date GCs and consequently a number of techniques have been employed to overcome the age-metallicity degeneracy \citep[e.g. ][]{1994ApJS...95..107W}.

Analysis of absorption line indices from spectroscopy is one such technique and most of the few hundred or so GCs with good data are found to be as old as Galactic GCs \citep[see summary of observations in ][]{2006ARA&A..44..193B}.  Unfortunately, the lengthy telescope time required for such analysis is costly, hence the number of GCs studied in this manner remains remarkably low.  A notable exception is the recent work of \citet{2008MNRAS.386.1443B}, who have obtained spectroscopic ages and metallicities for $\sim150$ GCs in the massive elliptical galaxy, NGC~5128.  Existing spectroscopic studies are also biased towards the brightest GCs, an important caveat that should be re-emphasised given the discovery of a GC colour-magnitude trend \citep{2006ApJ...636...90H,2006AJ....132.2333S,2006AJ....132.1593S,2006ApJ...653..193M}.  On the other hand, the well-studied Milky Way GC system exhibits no such trend with GC luminosity.

GC colour data can also help overcome the age-metallicity degeneracy.  This involves a comparison between observational colour-colour data for individual GCs and theoretical stellar population model predictions.  Optical to infrared (IR) colours are essential for this work, as the degeneracy among optical colours is severe for ages $\ga2$Gyr (e.g. Lee et al. 2007a).  From analysis of optical and near-IR photometry, GC system age substructure has been reported in NGC~1316 \citep{2001MNRAS.328..237G}, NGC~4365 (\citealt{2002A&A...391..453P,2005ApJ...634L..41K}; however c.f. \citealt{2005A&A...443..413L,2005AJ....129.2643B}) and NGC~5846 \citep{2003A&A...405..487H}.  In most instances, a subpopulation of ``intermediate-age'' GCs ($\sim2-8$ Gyr) that make up a small fraction ($\sim15\%$) of the total GC system is claimed.  A notable exception is the GC system of NGC~4365, which might contain $\sim40-80\%$ of these younger GCs (Puzia et al. 2002; though cf. Larsen et al. 2005 \& Brodie et al. 2005).  The systems for which no age substructure was reported using near-IR photometry include:  M87, NGC~4478 (Kissler-Patig, Brodie \& Minniti 2002), NGC~3115 \citep{2002A&A...391..453P}, NGC~4594, NGC~3585 and NGC~4472 (Hempel et al. 2007).  However, the data samples for each of these systems are typically small:   on order of a few tens of GCs per system.

Optical to IR colour data is also well-suited for characterising GC metallicity distributions, hence can help constrain the chemical properties of galaxies at early epochs.  Such colours span very large dynamical ranges and allow more precise metallicity measurements.  For example, 0.1 dex in metallicity corresponds to $\sim0.03$ mag in B--R while 0.1 dex is $\sim0.07$ mag in B--K (according to the preliminary models of Charlot \& Bruzual 2008 priv. comm.).  The IR is also insensitive to hot horizontal branch stars.  This is a important point considering the cautionary work of Yoon, Yi \& Lee (2006) who suggested GC colour bimodality might entirely result from a rapid transition between blue and red horizontal branches at intermediate GC metallicities.

The above issues are addressed here with an observational study of the Spitzer Space Telescope \citep{2004ApJS..154....1W} IR Array Camera \citep[IRAC; ][]{2004ApJS..154...10F} mid-IR properties of GC systems in the NGC~4594 and NGC~5128 galaxies.  NGC~4594, the ``Sombrero galaxy'' is classified as an Sa-type spiral, but has a galaxy stellar mass-normalised number of blue GCs \citep[T$_{blue}$; ][]{1993MNRAS.264..611Z} comparable to massive, early-type galaxies (Spitler et al. 2008).  With an estimated $1900$ GCs \citep{2004AJ....127..302R}, the NGC~4594 galaxy perhaps hosts the largest GC system within the local volume.  Recent Hubble Space Telescope (HST) Advanced Camera for Surveys (ACS) imaging revealed bi-modality in the optical colours of the GC system \citep{2006AJ....132.1593S}, confirming previous work in the optical:  \citet{1999AJ....118.1526G}, \citet{2001MNRAS.327.1116L}, and Rhode \& Zepf (2004).  Co-added spectra from 14 NGC~4594 GCs were found to be consistent with old ages by \citet{2002AJ....124..828L}.  Comparing V--I and V--K photometric data of 26 NGC~4594 GCs to stellar population models, Hempel et al. (2007) concluded these GCs are predominantly old.

NGC~5128, ``Centaurus A'', is a peculiar S0 galaxy whose GC system shows optical colour bimodality \citep{2001A&A...369..812R,2004ApJ...602..705P}.  Harris et al. (2006b) estimated the GC system contains $1550\pm390$ GCs.  A intermediate-{\it aged} subpopulation making up no more than 15\% of the GC system total was recently reported by \citet{2008MNRAS.386.1443B}.  In addition to providing spectroscopic evidence for the metal-poor and metal-rich GC subpopulations, \citet{2008MNRAS.386.1443B} also found a small, intermediate-{\it metallicity} GC subpopulation.

In the following, the distances to NGC~4594 and NGC~5128 galaxies are taken to be m$-$M$=29.77\pm0.03$ or 9.0 Mpc \citep{2006AJ....132.1593S} and m$-$M$=27.74\pm0.14$ or 3.5 Mpc \citep{2007ApJ...654..186F}, respectively.

\section{Data}\label{data}

\subsection{Spitzer Data Reduction}\label{spitzerdata}

\begin{figure*}
\resizebox{0.7\hsize}{!}{\includegraphics{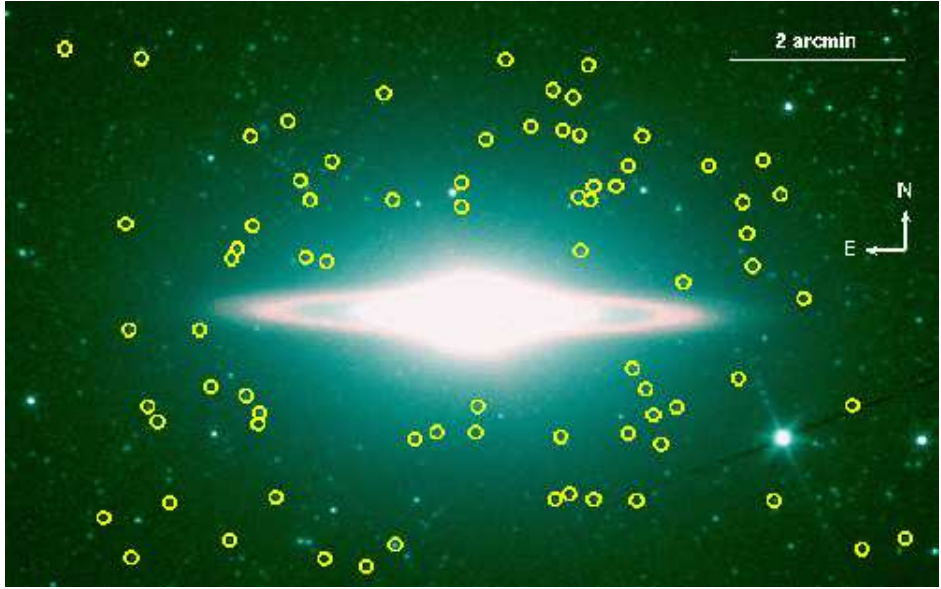}}
\caption{Spitzer IRAC \f-band image of the NGC~4594 galaxy.  GCs candidates from HST optical data (Spitler et al. 2006) that had counterparts on the IRAC images are highlighted with yellow circles.}\label{figspitzer}
\end{figure*}

\begin{figure*}
\resizebox{0.7\hsize}{!}{\includegraphics{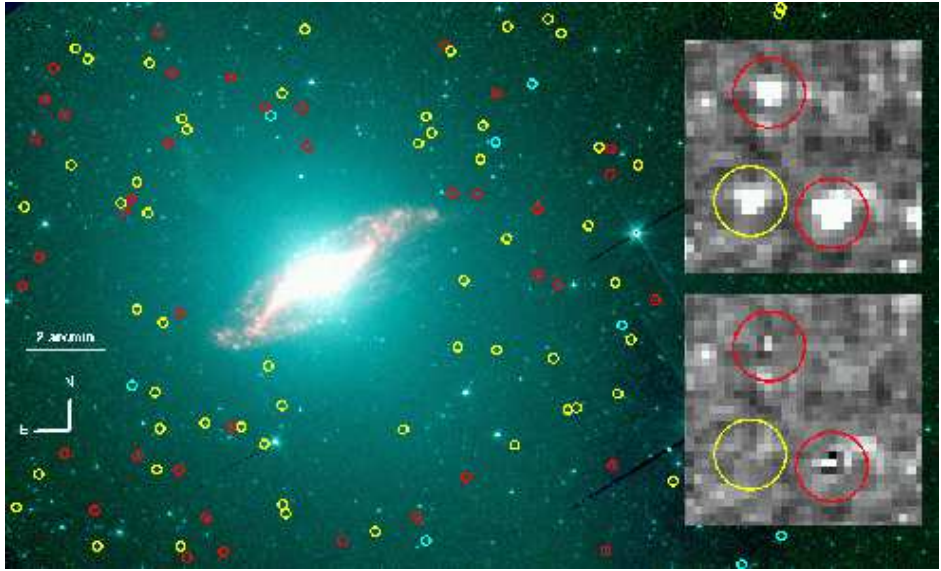}}
\caption{Spitzer IRAC colour images of the NGC~5128 galaxy.  Circle symbols represent the location of confirmed GCs from the catalogue of Woodley et al. (2007) that were successfully matched to objects in the Spitzer \f  imaging.  Red, cyan and yellow  circles represent GCs with spectroscopic old, intermediate to young and no ages, respectively according to the analysis of \citet{2008MNRAS.386.1443B}.  The lower zoomed inset shows residuals from PSF fits to the [3.6] image presented in the upper inset.}\label{figcena}
\end{figure*}

Figure~\ref{figspitzer} presents IRAC images of the NGC~4594 galaxy produced by the Spitzer Infrared Nearby Galaxies Survey (SINGS) legacy program \citep{2003PASP..115..928K}.  The IRAC third ([5.6]) and fourth ([8.0]) channel images (with filters centred on 5.6 and 8.0 microns, respectively) have an overwhelming background level and therefore GCs are not apparent in these images.  In contrast, IRAC channels one (\f) and two ([4.5]) show many sources over the entire field of view.  NGC~4594 observations were designed so that the central mosaic at least had $8\times30$s overlapping individual exposures and the \f field of view is $\sim13\arcmin\times25\arcmin$ ($\sim34$ kpc$\times65$ kpc).  The \f images were dithered by the SINGS group to increase the image resolution and astrometric precision.  The resultant pixel scale of $0.75\arcsec$pixel$^{-1}$ is still relatively large for this GC system, thus object crowding is problematic.

The NGC~5128 IRAC image set is associated with the Spitzer program \#101 (P.I. C. Lawrence) and was prepared for analysis with the standard Spitzer archive data pipeline.  The \f mosaic covers a $27\arcmin\times24\arcmin$ ($\sim29$ kpc$\times24$ kpc) region slightly offset from the galaxy centre and is shown in Figure~\ref{figcena}.  The mosaic average exposure time is approximately 72s (some locations on the mosaic have lower or higher coverage from the dithering pattern) and the pixel scale is $1.2\arcsec$pixel$^{-1}$.

\subsection{Spitzer Photometry}\label{datareduction}

The software tools used to extract IRAC photometry are packaged in the multi-purpose image analysis software application \citep[{\sc MOPEX}; ][]{2005PASP..117.1113M} and the specific method employed was that recommended in the {\sc MOPEX} documentation.  Object detection and photometry was carried out on the \f and [4.5] images after the diffuse galaxy light was removed by the subtraction of a median-filter image.  A local object detection threshold of $4\sigma$ was used.  To address object crowding, the apparent magnitude of each object was determined by fitting a point spread function (PSF) to its profile.  An empirical PSF for each IRAC dataset was constructed with {\sc MOPEX} using bright, isolated objects.  The PSF full-width at half-maximums are 1.7 (2.0\arcmin) and 2.1 (1.6\arcmin) pixels for the NGC~5128 and NGC~4594 \f images, respectively.  As demonstrated in Figure~\ref{figcena}, the PSF fitting residuals are minimal, suggesting the photometric method is robust.  Fluxes from individual objects were calibrated to the Vega magnitude system using the photometric zeropoints from \citet{2005PASP..117..978R}.  Apparent magnitudes were not corrected for Galactic extinction because dust has a negligible impact in this wavelength regime (A$_L \approx$ A$_{\f} \la 0.02$ mag).

 While the majority of the objects detected in the \f-band images are also found in the [4.5] images, the magnitude difference between the [3.6] and [4.5] bands are essentially negligible for old GCs \citep[e.g. ][]{2003MNRAS.344.1000B}.  For such stellar populations, if the [3.6] and [4.5] photometric data had shown comparable measurement uncertainties, the two bands might be combined to increase the accuracy of their mid-IR photometric measure.  In the present case, the formal [4.5]-band photometric uncertainties are twice those of the \f -band, thus the subsequent analysis only makes use of the \f -band data.

\begin{figure}
\resizebox{1\hsize}{!}{\includegraphics{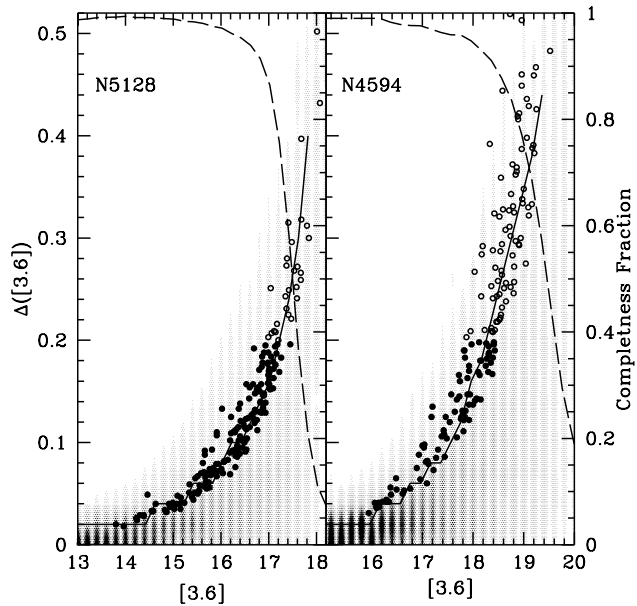}}
\caption{Profile fitting completeness results for NGC~5128 ({\it left}) and NGC~4594 ({\it right}).  For each panel, the left hand y-axis is the \f-band uncertainty and the right hand y-axis is the completeness fraction.  The residuals from the tests (i.e. the absolute value of the difference between the input and output magnitudes) are shown as the shaded region with the $1\sigma$ residual level marked as a solid line.  Green, solid circles represent objects analysed further ($\Delta$(\f)$<0.2$ mag), while the open circles were not used.  The dashed line corresponds to the completeness fraction estimated from the tests.}\label{figcomptest}
\end{figure}

To test the robustness of the Spitzer photometry extraction method, images containing artificial objects of known magnitudes were produced and their magnitudes were extracted using the same method described above.  In detail, for each 0.2 mag. interval over the observed magnitude range of the GCs, an artificial field from the original images field was populated with 4000 artifical objects.  The same detection and PSF-fitting method were applied to these artifical fields and the results are summarised in Figure~\ref{figcomptest}.  From the residuals between the input and measured magnitudes, it is apparent the formal photometric errors estimated by {\it MOPEX} were slightly underestimated.  The formal photometric errors are therefore taken as the $1\sigma$ level of the test residuals.

\subsection{Optical Photometry}\label{opticaldata}

NGC~4594 GC optical photometry was extracted from a HST ACS image mosaic (B,V, \& R bands) by \citet{2006AJ....132.1593S}.  The ACS imaging covers a $\sim10\arcmin \times 7\arcmin$ ($\sim26$ kpc$\times18$ kpc) region, centred on NGC~4594.  Optical photometry for the NGC~5128 GCs is from the Woodley et al. (2007) catalogue, which is largely drawn from the work of Peng et al. (2004).   A subset of the NGC~5128 GCs have spectroscopic ages and metallicities measured by \citet{2008MNRAS.386.1443B}.  In both optical datasets, corrections for Galactic extinction are from the DIRBE dust maps \citep{1998ApJ...500..525S}.

NGC~4594 GC photometry was derived in \citet{2006AJ....132.1593S} using a standard aperture photometry technique.  They applied a single aperture correction to account for light falling beyond the extraction aperture (5 ACS pixels or $0.25\arcsec$) and within a radius of 10 ACS pixels ($0.50\arcsec$).  A constant aperture correction for the entire dataset may not accurately represent the GC photometry.  More specifically, a trend is observed where the more extended objects show $\sim0.2$ mag. more flux than the most compact GC in the 5 to 10 pixel annulus.  The effect is roughly the same in each of the B, V and R bands thus the effect was negligible in the Spitler et al. (2006) work because it will cancel when optical colours are produced.  The cancelling occurs because the pixel scales of the ACS images are identical and filter-dependent PSF variations are small.  However, when ACS and Spitzer photometry are combined, the aperture corrections applied must be performed properly in an absolute sense to avoid introducing scatter in the ACS-Spitzer colours.  Such scatter results from the different pixel scales of the instruments:  GCs are resolved in the ACS images, but not in the IRAC images.  To mitigate this effect, the NGC~4594 optical photometry was re-derived with an extraction aperture of 10 pixels and adopted throughout the following.

\subsection{Optical/Spitzer Cross-matching}\label{match}

Cross-matching between the optical GC catalogues and Spitzer objects was carried out using a matching threshold of $1.08\arcsec$ in both the NGC~4594 and the NGC~5128 images.  In the NGC~4594 dataset, to decrease the chances of a nearby (projected) object effecting Spitzer flux measurements, all GC candidates with at least one nearby ($<1.6\arcsec$) object in the ACS images of a comparable magnitude (within +2.5 B-band magnitudes of the GC candidate) were culled from further analysis.  A total of 19 NGC~4594 GC candidates were not considered for this reason.  NGC~5128 GCs were not subject to this procedure because optical imaging is not publicly available.  A single NGC~4594 GC candidate shows a significant excesses of [3.6] light expected for GCs (Spitler et al. [2006] ID\#538; with V $=22.98$ and [3.6] $=17.81$).  This GC candidate was removed from further analysis as a possible contaminate, although flux confusion from an object undetected on the ACS imaging (e.g. a distant red galaxy) cannot be ruled out.

A photometric error selection of $\Delta($\f$)\le0.2$ mag was applied.  This roughly corresponds to M$_{\f}^{N5128}\la -10.2$ and M$_{\f}^{N4594}\la -11.4$ mag.  A small fraction of the NGC~5128 optical photometry is very uncertain, therefore an additional selection of $\Delta($\br $)\le0.1$ mag. was applied.

After this selection, 76 and 184 GC candidates with both optical and Spitzer photometry remained in the NGC~4594 and NGC~5128 datasets, respectively.  Of the matched NGC~5128 GCs with good spectroscopic age measurements (91) from \citet{2008MNRAS.386.1443B}, 11 or $\sim 12\%$ show ages less than 8 Gyr.  All [3.6] photometry is available from the first author and is posted on the CDS.

A very luminous, compact object (V $=17.46$) that is well-resolved on the HST ACS imaging, was also detected on the [3.6]-band imaging and has \f $=14.19$.  A radial velocity measurement from the spectroscopic sample of Hau et al. (2008, in prep.) suggests this object is associated with the NGC~4594 galaxy, hence is considered a ultra-compact dwarf galaxy (UCD) candidate.  It is included in the following analysis when appropriate.

\section{Results/Discussion}\label{results}

\subsection{Optical/Mid-IR Colour Distributions}\label{colourdists}

\subsubsection{Bimodality}\label{bimodal}

\begin{figure}
\resizebox{1\hsize}{!}{\includegraphics[angle=0]{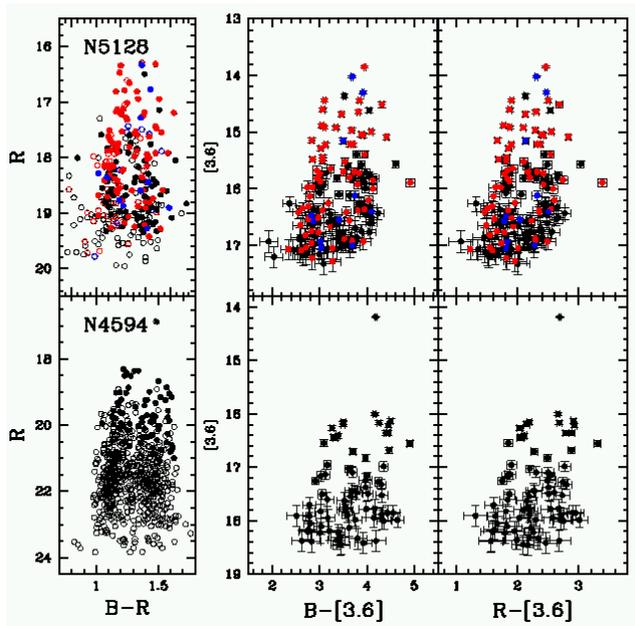}}
\caption{Colour-magnitude diagrams for NGC~5128 ({\it top}) and NGC~4594 ({\it bottom}) GCs.  A UCD associated with the NGC~4594 galaxy is given in the two lower diagrams at bright magnitudes.  Open circles in the two left plots have no corresponding match on the Spitzer imaging.  For NGC~5128, red, blue and black circles represent GCs with spectroscopic ages from \citet{2008MNRAS.386.1443B} with $\ge8$Gyr, $<8$Gyr and unknown ages, respectively.  GC colours show bimodal distributions.}\label{figcmd}
\end{figure}

\begin{figure}
\resizebox{1\hsize}{!}{\includegraphics{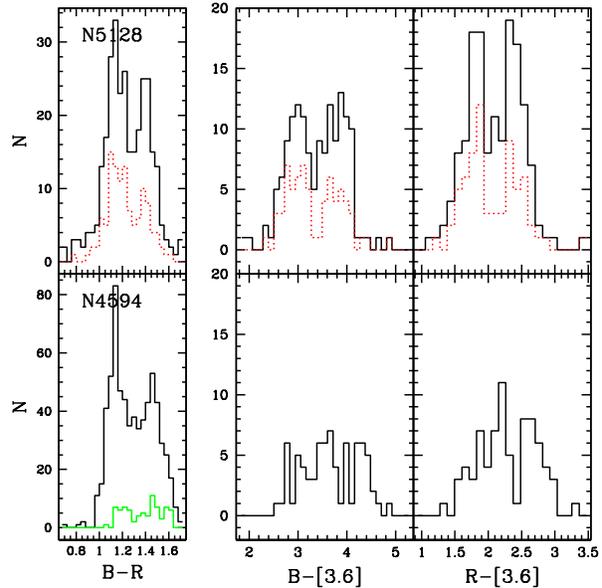}}
\caption{Same layout as Figure~\ref{figcmd}, with the corresponding colour histograms given.  Dashed red histograms show NGC~5128 GCs with old ($\ge8$Gyr) ages and the green histogram in the lower left plot shows the \br colour histogram of NGC~4594 GCs matched with Spitzer photometry.  Colour bimodality is most apparent in the optical, NGC~5128 \bc and NGC~5128 \rf distributions.}\label{fighist}
\end{figure}

GC colour-magnitude diagrams are presented in Figure~\ref{figcmd} for both GC systems.  Colour histograms are given in Figure~\ref{fighist}.  Bimodality in the optical to Spitzer colours is clearly evident in the Figures and the colour ``gap'' between the GC subpopulations is located at \bc $\sim3.4$ and \rf $\sim2.1$ mags.

The KMM algorithm of \citet{1994AJ....108.2348A} was used to quantify how likely these datasets are drawn from a bimodal distribution instead of a unimodal one.  When run on the NGC~5128 \bc data, the KMM output strongly supports colour bimodality at a highly significantly level (see Table~\ref{tabkmm}).  If a KMM analysis is performed only on GCs that are spectroscopically determined to be old, the statistical confidence for \bc bimodality remains strong.  

The KMM results support the idea that the NGC~4594 \bc distribution is well-represented as two Gaussian compared to a single Gaussian.  However, the KMM \bc colour peaks are $\sim0.5$ mag redder than what can be inferred from the colour-magnitude diagram in Figure~\ref{figcmd}.  Two features found by Spitler et al. (2006) in the NGC~4594 colour-magnitude parameter space likely influence the KMM results:  the existence of bright GCs with colours consistent with the subpopulation gap and blue GCs show a colour-magnitude trend (the ``blue tilt''; see \S\ref{tilt}), thus brighter blue GCs will tend to ``fill-in'' any gap between two colour subpopulations.  A more complete sampling of the GC luminosity function is required before the \bc KMM output is to be trusted. 

This analysis confirms GC colour bimodality previously found using optical photometry (see references in the Introduction).  It is apparent that the larger baseline of the optical to Spitzer colours helps identify the GC subpopulations more clearly than when only optical photometry is used.

\begin{table}
 \centering
  \caption{Results from a KMM analysis of the GC \br, \bc, and \rf distributions.  The p-stat value is the bimodality gauge produced by the KMM algorithm (see text).  The last two columns are the mean colour of each GC subpopulation estimated by KMM.  To prevent data outliers from influencing the KMM tests, the NGC~4594 colours were restricted to \bc $\le4.8$ and the NGC~5128 data to $2.4<$ \bc $\le 4.2$.  The \rf data is bounded by $1.3<$ \rf $<3.2$.  The heteroscedastic version of the KMM test is employed for the \bc and \rf colour distributions.}\label{tabkmm}
  \begin{tabular}{@{}lcrrrr@{}}
  \hline
GC Sample & Colour & Numb. & log(p-stat) & blue & red \\
 &  & of GCs & & peak & peak \\
 \hline
 \hline
\multicolumn{6}{l}{NGC~5128}\\ 
 \hline
 $\Delta\f\le0.2$ & \br &  284 & -7.5 & 1.14  & 1.41 \\
age $\ge8$        & \br &  113 & -3.5 & 1.15 & 1.41 \\
 $\Delta\f\le0.2$ & \bc&  140 & -5.7 & 2.96 & 3.78 \\
age $\ge8$        & \bc&  64  & -3.8 & 2.94 & 3.78 \\
 $\Delta\f\le0.2$ & \rf&  146 & -5.7 & 1.78 & 2.39 \\
age $\ge8$        & \rf&  66  & -2.8 & 1.77 & 2.40 \\
\hline
\multicolumn{6}{l}{NGC~4594}\\ 
 \hline
 $\Delta\f\le0.2$ & \br &  654 & -20.  & 1.16 & 1.46 \\
 $\Delta\f\le0.2$ & \bc &  75 & -2.0  & 3.43 & 4.34 \\
 $\Delta\f\le0.2$ & \rf &  75 & -0.4  & 1.72 & 2.41 \\
\hline
\end{tabular}
\end{table}

\subsubsection{The Blue Tilt}\label{tilt}

\begin{figure}
\resizebox{1\hsize}{!}{\includegraphics[angle=0]{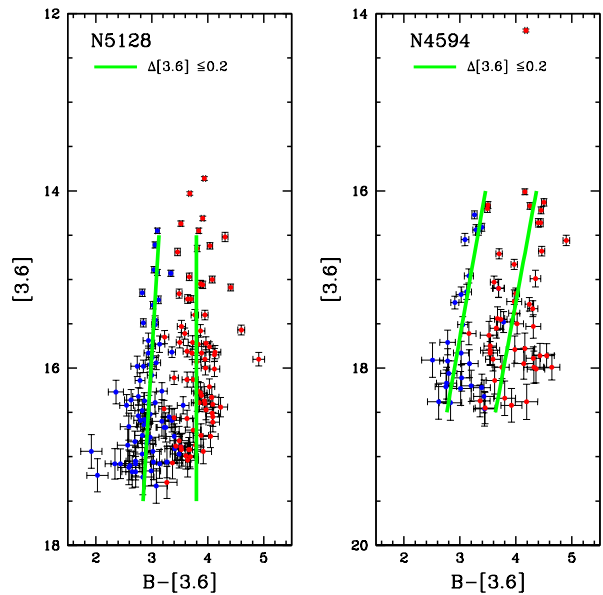}}
\caption{Spitzer to optical colour-magnitude diagrams.  Blue and red points correspond to objects with \br $\le1.3$ and \br $>1.3$, respectively.  This split on \br corresponds to the GC subpopulation division used by Spitler et al. (2006).  Lines fitted to the subpopulation colour-magnitude data are given.  A blue GC colour-magnitude trend (a ``blue tilt'') is found in the NGC~5128 data, while incomplete luminosity coverage of the NGC~4594 data limits such analysis.}\label{figcmdsubpop}
\end{figure}

Spitler et al. (2006) discovered a trend among the blue NGC~4594 GCs, where the colours become redder with increasing GC luminosity.  Harris et al. (2006a) pointed out a similar trend is apparent in a figure of Harris et al. (2004) containing NGC~5128 photometry.  To explore this so-called ``blue tilt'' (Strader et al. 2006) in the optical to Spitzer parameter space, the GC systems was divided into a blue and red subpopulation at \bc $\le3.4$ and \bc $>3.4$, respectively.  Each subpopulation was then fitted with a line to the \bc versus \f information (weighted by the formal photometric uncertainties).  Parameters are in Table~\ref{tabtilt} and the fits themselves shown in Figure~\ref{figcmdsubpop}.

The linear fits to the NGC~5128 GC subpopulation data confirm the qualitative assessment of existing optical photometry by Harris et al. (2006a) and suggest the blue subpopulation shows a very significant ($\sim9\sigma$) \bc versus \f colour-magnitude gradient, while the red subpopulation shows no such trend.  Fitting only to old NGC~5128 GCs yields a similar outcome.  Converting the NGC~5128 to a metallicity--mass proportionality with the empirical colour-metallicity transformations described in Sec.~\ref{spitzermetal}, the NGC~5128 blue GCs show $Z\propto M^{0.19}$.  The strength of this effect is among the lowest reported thus far.  The approximate mass range covered by the Spitzer data is $\sim5\times10^{5}$ to $10^{7} M_{\odot}$.

Fits to the NGC~4594 \f versus \bc data suggests {\it both} GC subpopulations show a formally significant ($\sim10\sigma$) colour-magnitude trend (see the fits in Fig.~\ref{figcmdsubpop}).  A red subpopulation trend was not found in the optical analysis by Spitler et al. (2006).  A better sampling of the GC luminosity function (which currently only covers $\sim10^{6}$ to $10^{7} M_{\odot}$) is likely required before such trends are constrained with Spitzer mid-IR photometry.

The cause of blue tilt is not currently understood though a number of ideas have been proposed (see discussion in Strader et al. 2006; Harris et al. 2006a; Mieske et al. 2006; Spitler et al. 2006).  The present work rules out age effects as the cause of the NGC~5128 blue tilt.  If the trend is due to increasing metallicity with luminosity, an analogous trend for red GCs should be a factor $\sim10$ times shallower than the gradient observed among the blue GC subpopulation (Spitler et al. 2006), hence a good reason for it not being observed if it is indeed present.  Determining whether or not a metallicity-induced red tilt actually exists would clearly be aided with the long baseline offered by Spitzer to optical colours.

\begin{table}
 \centering
  \caption{Linear fits to GC subpopulations.  The NGC~4594 data were restricted to \bc $\le4.8$ and the NGC~5128 to $2.4<$ \bc $\le 4.3$ before the fits were performed.}\label{tabtilt}
  \begin{tabular}{@{}llrrr@{}}
  \hline
GC & GC & N$_{GC}$ & \multicolumn{2}{c}{ (\bc) $=\alpha$ \f $+ \beta$ }\\ 
Sample  & Subpop & & $\alpha$ & $\beta$ \\
 \hline
 \hline
\multicolumn{5}{l}{NGC~5128}\\ 
 \hline
 $\Delta\f\le0.2$ & red  &  73 & $+0.001\pm0.008$  & $3.79\pm0.12$ \\
 $\Delta\f\le0.2$ & blue &  68 & $-0.094\pm0.013$  & $4.49\pm0.20$ \\
old		  & red  &  27 & $-0.024\pm0.048$  & $4.16\pm0.76$ \\
old		  & blue &  37 & $-0.095\pm0.042$  & $4.47\pm0.68$ \\
\hline
\multicolumn{5}{l}{NGC~4594}\\ 
 \hline
 $\Delta\f\le0.2$ & red  &  46 & $-0.293\pm0.020$  & $9.06\pm0.34$ \\
 $\Delta\f\le0.2$ & blue &  28 & $-0.277\pm0.025$  & $7.89\pm0.42$ \\
\hline
\end{tabular}
\end{table}

\subsection{Optical-Spitzer/Metallicity Relationships}\label{spitzermetal}

\begin{figure*}
\resizebox{1\hsize}{!}{\includegraphics[angle=0]{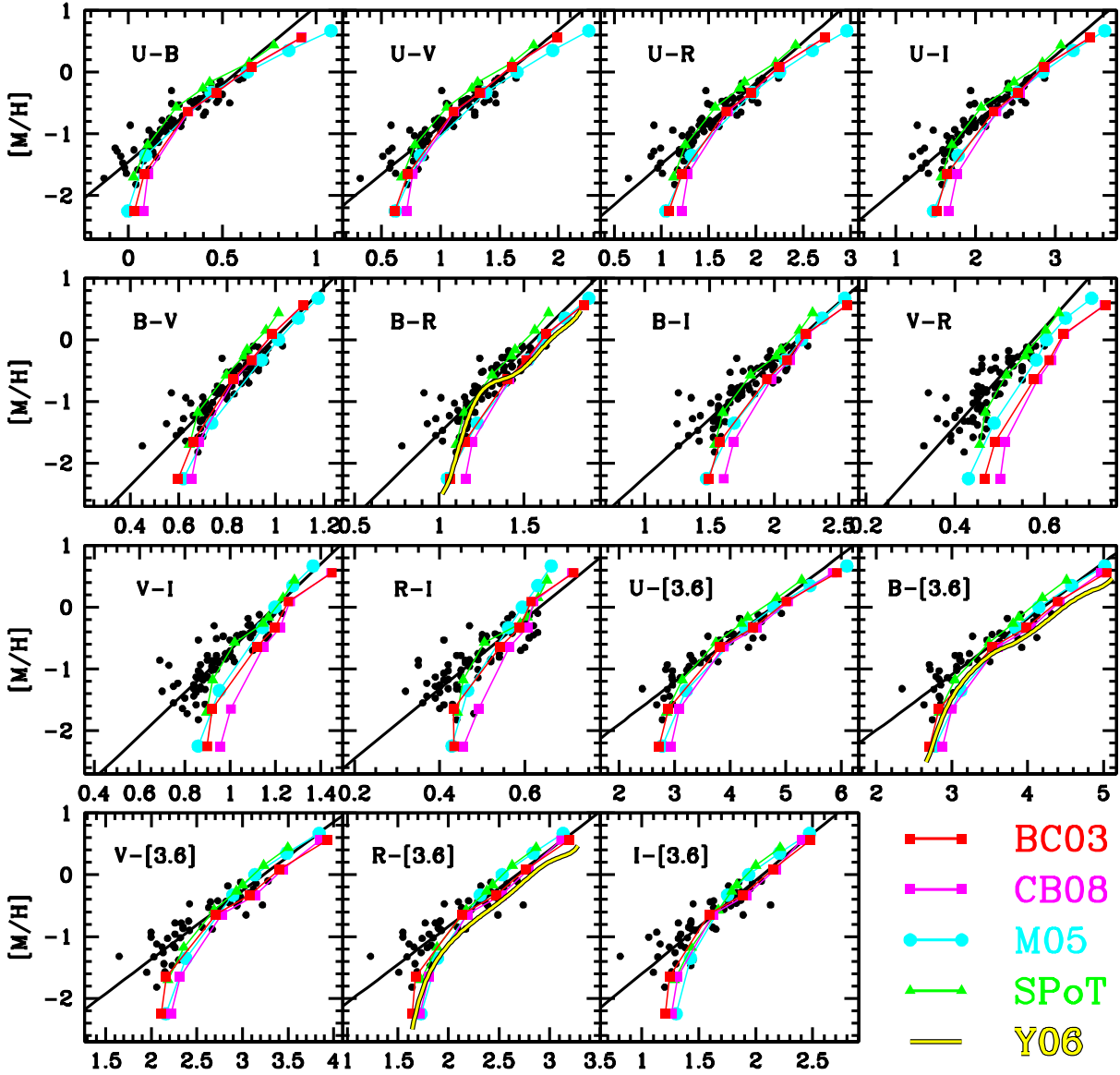}}
\caption{Colour-metallicity diagrams of NGC~5128 GCs with ages $\ge 8$ Gyr and metallicity uncertainties of $<0.15$ dex.  NGC~5128 metallicities ([M/H]; scale is from Harris 1996) are from \citet{2008MNRAS.386.1443B}.  Black lines are the best fitted linear relationships to the data using a least-squares fit (clipping outliers, as appropriate).  The parameters of these fits are presented in Table~\ref{tabspitmetal}.  Predictions from five stellar population models are shown.}\label{figspitmetal}
\end{figure*}

\begin{table}
 \centering
  \caption{Empirical NGC~5128 GC optical to mid-IR colour-metallicity relations.  Valid for use with old ($>8$ Gyr) single stellar populations spanning a metallicity range of $-1.8 \la$ [M/H] $\la+0.0$.  Metallicity scale is that of the Harris (1996) Galactic GC catalogue.  Optical colours are corrected for Galactic dust extinction using the maps of Schlegel et al. (1998).}\label{tabspitmetal}
  \begin{tabular}{@{}lrrrrrr@{}}
  \hline
& \multicolumn{3}{c}{[M/H]$=\alpha(X)_0+\beta$} \\
   $X$ & $\alpha$ & $\beta$ & RMS \\
 \hline
U--B    & $2.51\pm0.14$ & $-1.45\pm0.05$ & $0.13$ \\
U--V    & $1.56\pm0.09$ & $-2.43\pm0.07$ & $0.16$ \\
U--R    & $1.32\pm0.08$ & $-2.81\pm0.08$ & $0.18$ \\
U--I    & $1.08\pm0.07$ & $-2.98\pm0.09$ & $0.20$ \\
B--V    & $4.00\pm0.27$ & $-3.96\pm0.12$ & $0.14$ \\
B--R    & $2.71\pm0.20$ & $-4.23\pm0.14$ & $0.16$ \\
B--I    & $1.88\pm0.13$ & $-4.11\pm0.13$ & $0.18$ \\
V--R    & $8.05\pm0.81$ & $-4.63\pm0.19$ & $0.16$ \\
V--I    & $3.48\pm0.29$ & $-4.17\pm0.15$ & $0.16$ \\
R--I    & $5.65\pm0.52$ & $-3.59\pm0.15$ & $0.16$ \\
U--[3.6]        & $0.69\pm0.05$ & $-3.27\pm0.11$ & $0.23$ \\
B--[3.6]        & $0.91\pm0.07$ & $-3.80\pm0.14$ & $0.22$ \\
V--[3.6]        & $1.11\pm0.09$ & $-3.60\pm0.14$ & $0.21$ \\
R--[3.6]        & $1.24\pm0.11$ & $-3.31\pm0.14$ & $0.21$ \\
I--[3.6]        & $1.51\pm0.15$ & $-3.12\pm0.14$ & $0.21$ \\
C--[3.6]        & $0.76\pm0.06$ & $-3.88\pm0.15$ & $0.25$ \\
M--[3.6]        & $1.04\pm0.11$ & $-4.07\pm0.20$ & $0.25$ \\
T1--[3.6]       & $1.23\pm0.14$ & $-3.73\pm0.19$ & $0.24$ \\
C--M    & $2.30\pm0.16$ & $-2.82\pm0.10$ & $0.16$ \\
C--T1   & $1.71\pm0.13$ & $-3.63\pm0.13$ & $0.18$ \\
M--T1   & $5.25\pm0.95$ & $-4.86\pm0.27$ & $0.23$ \\
\hline
\end{tabular}
\end{table}

Figure~\ref{figspitmetal} provides all Johnson/Cousins optical to optical and optical to Spitzer colour permutations against GC metallicity for the $\ge8$ Gyr NGC~5128 GC data.  The colour-metallicity data are well-represented with linear fits, which are presented in the same Figure and detailed in Table~\ref{tabspitmetal}.  These relationships are only valid for stellar population with ages $\ga8$ Gyr and metallicities of $-1.8 \la$ [M/H] $\la+0.0$.  Individual NGC~5128 GC metallicities were derived in \citet{2008MNRAS.386.1443B} and the technique employed is briefly summarised here.  For each GC, \citet{2008MNRAS.386.1443B} took every Lick index measured and transformed it into a metallicity value using empirical relationships derived from Galactic GC data.  The Tukey biweight of the all Lick index metallicity values is taken as the final metallicity ([M/H]) for a particular GC.  The [M/H] scale is that of Harris (1996) for the Galactic GC system.  Individual NGC~5128 ages were estimated with fits to the Lee \& Worthey (2005) SSP models.

The optical to [3.6] colours show a stronger correlation with metallicity compared to optical-optical colours (see formal significance of the fitted slopes in Table~\ref{tabspitmetal}).  To what degree this reflects the possibility that a linear fit is more appropriate for optical to [3.6] colours or else the fact that \f is relatively insensitive to second parameters (such as age and horizontal branch morphology) is unknown.  Certainly the latter point has some influence and the larger dynamical range provided by optical to [3.6] colours makes them a preferred metallicity proxy over optical to optical colours.

Certain SSP models (e.g. Bruzual \& Charlot 2003 hereafter BC03; Charlot \& Bruzual 2008 priv. comm., CB08; Maraston 2005, M05; S. Yoon 2006 priv. comm. , Y06) provide optical to Spitzer colour predictions for a SSP of a given age and metallicity.  As GCs are {\it single} stellar populations whose stars show a single age and metallicity, they provide an ideal metric to contrast different model predictions for old stellar populations (e.g. Proctor, Forbes, \& Beasley 2004; M05; Mendel et al. 2007; Cantiello \& Blakeslee 2007; Cohen et al. 2007; Pessev et al. 2008).  Figure~\ref{figspitmetal} provides the 12 Gyr colour-metallicity prediction for five SSP models:  BC03, CB08 (preliminary update to BC03), M05 (blue horizontal branch), Raimondo et al. (2005; SPoT), and Y06.  Note for the SPoT models, the L-band predictions are used in place of the [3.6]-band, which are currently not available.  For Y06, 13 Gyr models are used to best illustrate their predicted non-linear structure due to changing horizontal branch (HB) morphology (see Sec.~\ref{bimodality}).

It is apparent in Figure~\ref{figspitmetal} that all model colour predictions are roughly consistent with each other.  When the models are compared to the observed NGC~5128 GC colours, it is apparent at [M/H] $\la-1.5$ the models predict steeper [M/H]-colour gradients.  This is especially pronounced in colours containing redder optical wavelengths such as $R-$, $I-$, and [3.6]--band.  The apparent internal consistency between the models, whose treatment of HB stars differs (e.g. blue vs. red HBs), argues against HB morphology as the primary cause for the discrepancy between the data and models.  The BC03 and CB08 models tend to predict redder optical-[3.6] colours for a given metallicity at [M/H] $\ga-0.5$.

The empirical linear colour-metallicity relations provided in Table~\ref{tabspitmetal} should prove useful when the observed photometric uncertainties are comparable to the uncertainties of the NGC~5128 GC photometry.  An attempt to better constrain the SSP models with the observations is made in Section~\ref{models}, where the \rf colour is used as a model-independent proxy for metallicity.

\subsubsection{Does GC Colour Bimodality Imply Metallicity Bimodality?}\label{bimodality}

Yoon et al. (2006; see also Richtler 2006) recently demonstrated that the interpretation of a bimodal optical colour distribution as evidence for two subpopulations of GCs depends critically on the relationship employed to translate colour into metallicity.  According to their SSP models (Y06), the transition from a blue to red horizontal branch (HB) at intermediate metallicities causes a change in slope of this relation.  This change in slope means that an observed bimodal colour distribution may in reality be resultant from a single, unimodal metallicity distribution.

To determine how much of an effect HB stars have on the observed GC colour bimodality, colours derived from wavebands that are insensitive to HB stars (e.g. IR) can be compared to those colours that are (e.g. B-band).  Indeed, simulations by Cantiello \& Blakeslee (2007) of the SPoT models show that certain optical/near-IR colours (e.g. V--H and V--K) are preferred for circumventing any biases caused by any non-linear structure in the colour-metallicity relationship.  They also demonstrate that the non-linear structure is not consistent between the models they examined (they compared 5 different SSP models), thus any effect like that described in Yoon et al. (2006) is strongly dependant on the particular colour-metallicity relationship adopted.

S. Yoon (2006, priv. com.) kindly provided [M/H] predictions from the same SSP models used in Yoon et al. (2006) for three colours (\br, \bc, and \rf), which are shown in Figure~\ref{figspitmetal}.  The inflection point from a changing horizontal branch type at [M/H] $\sim-0.8$, is apparent in their \br model predictions.  The predicted [M/H] versus \rf relationship of Y06 in Figure~\ref{figspitmetal} is approximately linear, suggesting horizontal branch stars are no longer playing an important role at these wavelengths.  S. Yoon (2006, priv. com.) included a cautionary note that, in general, mid-IR stellar libraries are poorly populated, thus any model prediction at these wavelength are not well constrained.  Nevertheless, the relative insensitivity of \rf to HB stars predicted by Y06 makes them a useful colour combination to test the scenario of Yoon et al. (2006).

Analysis in Sec.~\ref{bimodal} of the old NGC~5128 GC \rf data provides strong evidence for colour bimodality.  Since this colour is insensitive to horizontal branch stars, these observations suggest colour bimodality is likely not caused by a variable horizontal branch morphologies, as suggested by Yoon et al. (2006).  Given the old ages of these GCs, the NGC~5128 colour distribution implies a bimodal metallicity distribution.  This is in agreement with the results of \citet{2008MNRAS.386.1443B} using spectroscopic data of NGC~5128 GCs.

Although the NGC~4594 optical to mid-IR colour data are not strongly bimodal, there is a good reason why this is expected.  In Figure~\ref{figcmd}, it is apparent that the NGC~4594 GCs with Spitzer photometry only cover $\sim2$ magnitudes of the GC luminosity function.  As the brightest, blue GCs are biased towards redder colours, they effectively ``fill-in'' the subpopulation gap, hence making any inherent bimodality more difficult to detect.  The NGC~5128 data samples almost twice the magnitude range, hence the blurring of the subpopulation data at high luminosities is less critical for this analysis.

Kundu \& Zepf (2007) concluded the observed bimodal I--H colour distribution of very bright GCs in the giant elliptical M87 as evidence for a bimodal metallicity distribution with a similar argument to that used here.  Strader, Beasley \& Brodie (2007; also Beasley et al. 2008) tackled the GC metallicity bimodality question without the use of theoretical models and employed empirical relationships they derived between metallicity and absorption line indices of Galactic GCs.  Although their technique did not explicitly account for variable HB morphology effects, they provide compelling evidence that luminous GCs in the giant elliptical NGC~4472 are drawn from a bimodal metallicity distribution.  

The optical to Spitzer colours produced here provide strong evidence for GC system {\it metallicity} bimodality in NGC~5128, and support recent results for bimodal GC metallicity distributions in certain giant elliptical galaxies.

\subsection{Colour-Colour Diagrams}\label{colourcolour}

\subsubsection{SSP Model Constraints}\label{models}

\begin{figure*}
\resizebox{1\hsize}{!}{\includegraphics[angle=0]{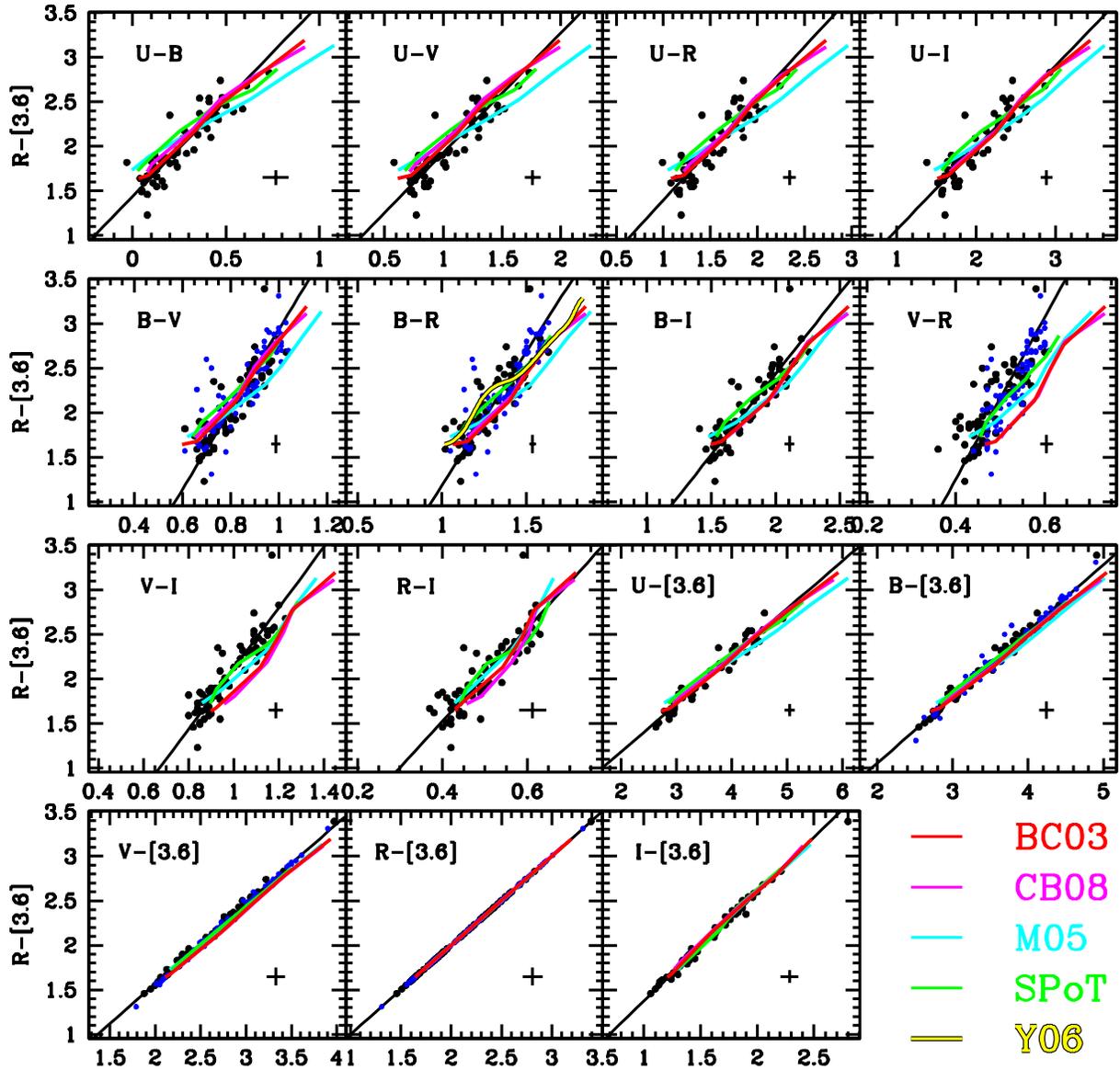}}
\caption{Colour-colour diagrams of NGC~5128 GCs with ages $\ge 8$ Gyr (black circles) and NGC~4594 GCs (blue circles).  The X-axis colour label is given within each plot.  Only data with X-axis colour errors of $<0.1$ mag are shown.  SSP models are as in Figure~\ref{figspitmetal}.  Lines are fits to the GC observations with:  C $=\alpha$(\rf)$ +\beta$, where C is the colour on the X-axis.  Mean photometric errors of the GC data are illustrated with the crosses in each diagram.}\label{figspitcolour}
\end{figure*}

The use of optical to Spitzer colours to test the colour predictions from different SSP models has certain advantages over metallicity.  The term ``metallicity'' is an ill-defined quantity and can vary between different SSP models (see discussion in Beasley et al. 2008).  Colours, on the other hand, are straight-forward to compute and therefore should be directly comparable between SSP models.  Optical to Spitzer colours are very good metallicity proxies and provide a higher precision due to increased dynamical ranges (\S\ref{spitzermetal}), hence they provide an excellent quantity for comparing theoretical predictions.

Figure~\ref{figspitcolour} shows the same colours of Figure~\ref{figspitmetal} instead plotted against the \rf colour.  The \rf was chosen over e.g. U-[3.6], B-[3.6], etc., for the following reasons:  (1) the NGC~4594 data can be included in the analysis; (2) it is relevant to the age-substructure analysis of \S\ref{ages}; (3) R-[3.6] shows most significant statistical correlations with [M/H] (see Table~\ref{tabspitmetal}).  The following analysis holds true if \bc is used instead as the basis.  Model predictions for 12 Gyr ages are presented in Figure~\ref{figspitcolour}.  The mean spectroscopic age of the NGC~5128 blue and red GC subpopulations are 11.1 and 10.8 Gyr, respectively, each with a $1\sigma$ age spread of $\sim0.9$ Gyr \citep{2008MNRAS.386.1443B}.  Analysis of roughly two dozen NGC~4594 GCs suggests they have old ages (Larsen et al. 2002; Hempel et al. 2007).  The following analysis does not change if the NGC~4594 data is not used.

\begin{figure*}
\resizebox{1\hsize}{!}{\includegraphics[angle=0]{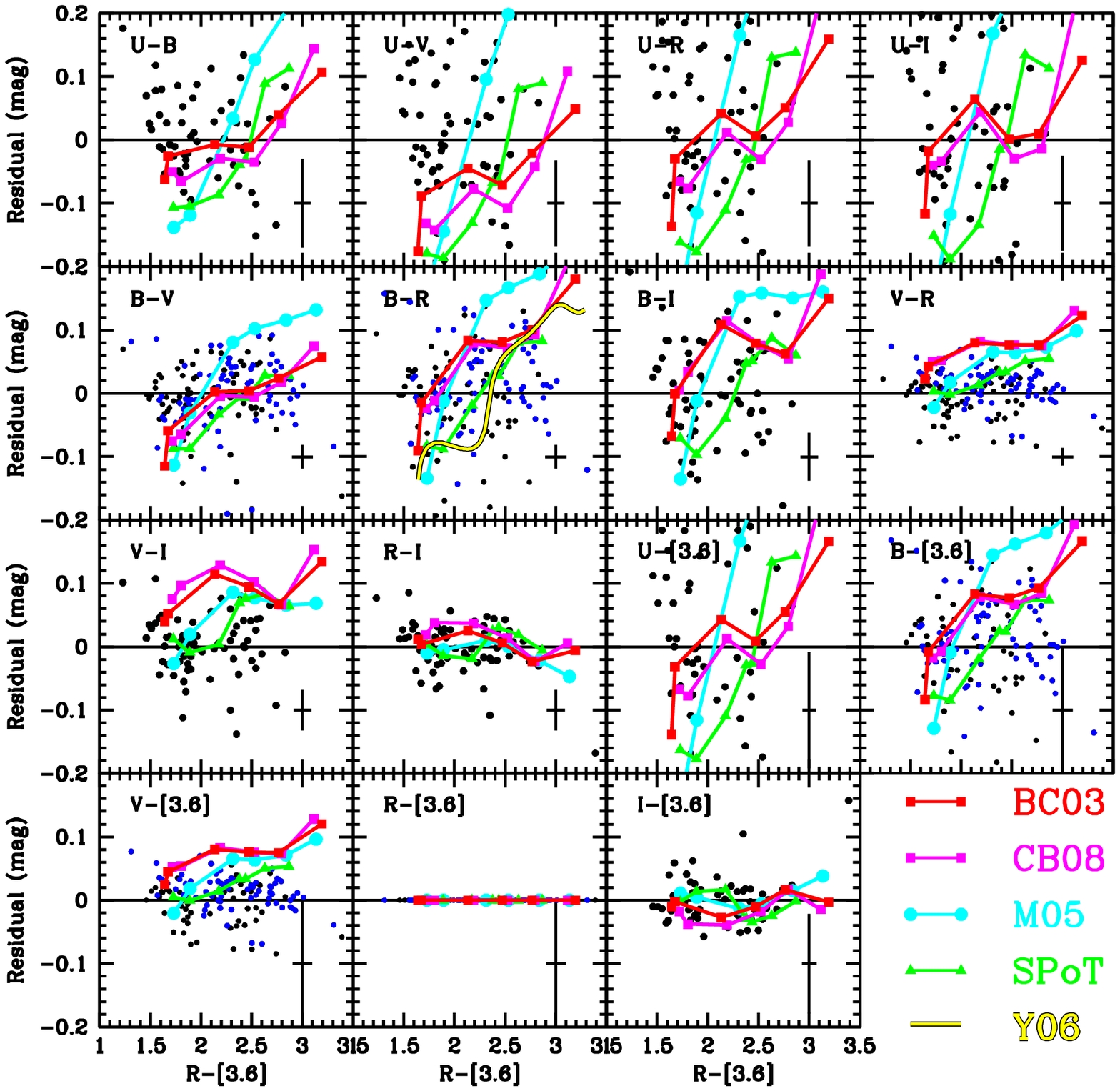}}
\caption{Data and model residuals from fits to the GC data in Figure~\ref{figspitcolour}.  Positive residuals correspond to colours redder than the fit to the GC data.  See text for discussion.}\label{figspitcolourres}
\end{figure*}

Overall the agreement between the model predictions and the data is very good.  In Figure~\ref{figspitcolour}, it is apparent that GC data generally reach lower values of \rf than predicted by the models, perhaps indicating the predicted [3.6] flux is underestimated at low metallicities.  To better facilitate a comparison, the data and model residuals from the fits to the data (shown in Figure~\ref{figspitcolour}) are plotted in Figure~\ref{figspitcolourres}.  Visually inspecting each optical wave-band in turn, the following observational constraints can be deduced from Figure~\ref{figspitcolourres}:

\begin{itemize}
\item $U-$band.  The BC03 and CB08 colour predictions best reflect the shape of the data.  The SPoT and especially M05 models predict bluer colours (more $U-$band flux) at low-metallicities and redder colours (less $U-$band flux) at high-metallicities for colours including this band.  CB08 provides the best match to the data at low-metallicities.
\item $B-$band.  The $B-$band model residuals resemble those of $U-$band.  M05 predicts redder colours (less $B-$band flux) at high metallicities compared to the other models and the data.  The other models actually show a similar residual structure to M05, although it is less pronounced.  Again, CB08 matches the data at the low metallicities.  At intermediate metallicities the SPoT models match the data well and show some resemblance to the Y06 models in the B--R diagram.
\item $V-$band.  For this filter, the M05 and SPoT models fair best, while BC03 and CB08 models evidently predict less V-band flux compared to the data at all metallicities.  This best illustrated in the V--R, V--I and V--[3.6] diagrams, but is also apparent when comparing the BC03 and CB08 predictions in the U--V diagram to U--B.
\item $R-$band and $I-$band.  The predictions and data appear to be in good agreement with each other, as demonstrated by the R--I and I--[3.6] diagrams.  If this is confirmed, it also suggests the data and model predictions at [3.6] are fairly consistent.
\end{itemize}

To conclude, optical to Spitzer colours are shown to provide a useful basis to help constrain SSP model predictions at optical wavelengths.  A comparison between the NGC~5128 and NGC~4594 GC data suggests SSP model predictions at bluer wavebands ($U$, $B$, and $V$) show the largest discrepancies with the GC data compared to the red bands ($R$ and $I$).

\subsubsection{Photometric Ages?}\label{ages}

\begin{figure*}
\resizebox{1\hsize}{!}{\includegraphics[angle=0]{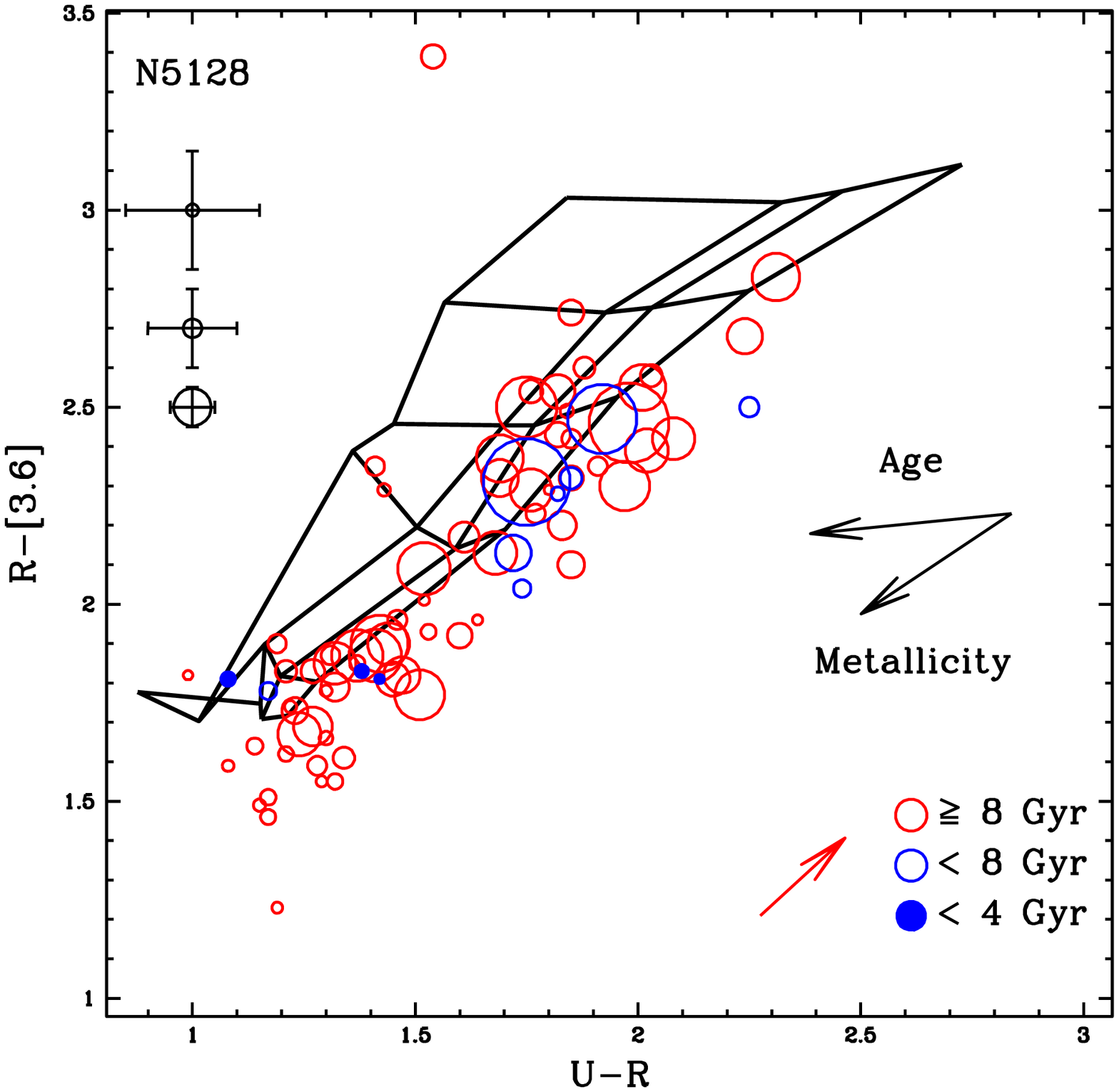}}
\caption{NGC~5128 GC \rf vs. U--R colour-colour diagram.  To highlight the most accurate photometry, the circles are inversely proportional to photometric uncertainties.  Blue and red circles correspond to GCs with spectroscopic ages of $<8$ and $\ge8$ Gyr, respectively.  GCs with ages $<4$ Gyr are shown as solid circles.  CB08 SSP model 2, 5, 7 and 12 Gyr predictions are presented and the arrows indicate the general directions in the models for {\it decreasing} age and metallicity.  A dust extinction arrow of A$_V=0.1$ mag. is given on the lower right.}\label{figurrch1}
\end{figure*}

GC colour-colour plots which include K-band photometry (V--I versus V--K is popular for GC work) have been used to look for age-substructure in GC systems (see references in the Introduction).   In theory, this technique works because an optical to optical colour (e.g. V--I) is slightly more sensitive to age, while an optical to K-band colour (e.g. V--K) is sensitive to metallicity.  The NGC~5128 GC spectroscopic analysis from \citet{2008MNRAS.386.1443B} enables the first large comparison between spectroscopic GC ages and those derived from optical and IR colour-colour diagrams.  Although the spectroscopic ages are not free from uncertainties, they better break the age-metallicity degeneracy (e.g. Worthey 1994).

Li \& Han (2008) investigated the relative sensitivities of all possible colour-colour pairs among the standard BVRIJHK$ugriz$ band-passes using the SSP of Bruzual \& Charlot (2003).  They concluded that the colour pairs, R--K versus $u$--R and $r-$K versus $u-r$, are most capable of partially breaking the age-metallicity degeneracy.  The closest colour combination in the NGC~5128 GC catalogue to the preferred pair found in Li \& Han (2008) is \rf versus U--R.  Analysis in Section~\ref{models} of Figure~\ref{figspitcolourres} suggests the CB08 provide the best match to old NGC~5128 GCs in this colour-colour parameter space.  GCs with reliable spectroscopic ages are presented in Figure~\ref{figurrch1} with the 12-, 7-, 5- and 2-Gyr SSP model predictions from CB08.

It is immediately apparent in Figure~\ref{figurrch1} that the young NGC~5128 GCs determined from spectroscopy show no systematic difference to the old GCs.  Even the youngest GCs ({\it c.f.} solid blue to open red circles) occupy the same parameter-space as the old GCs.  Relatively large photometric uncertainties and the inherent difficulty in breaking the age-metallicity degeneracy using colour-colour diagrams provide the most likely explanation for the discrepancy between the present findings and the \citet{2008MNRAS.386.1443B} spectroscopic ages.

If spectroscopic age information were not available for individual GCs, it might be tempting to tag a few of the more luminous metal-rich NGC~5128 GCs in Figure~\ref{figurrch1} as candidate young GCs because their colours coincide with a younger region of the SSP model grids.  Clearly, such a venture would prove erroneous once the spectroscopic ages were considered, suggesting any apparent age-substructure in colour-colour space {\it must} be investigated further with spectroscopic data.  Furthermore, had another model been employed for the colour-colour analysis (without the spectroscopic ages), the interpretation would have varied tremendously.  For example, the region where metal-rich GCs fall in Figure~\ref{figurrch1} matches with the $<5$ Gyr predictions of M05.  It is therefore apparent that accurately age dating GCs with colours alone is limited by the high level of systematic uncertainty in SSP models.

\begin{figure*}
\resizebox{1\hsize}{!}{\includegraphics[angle=0]{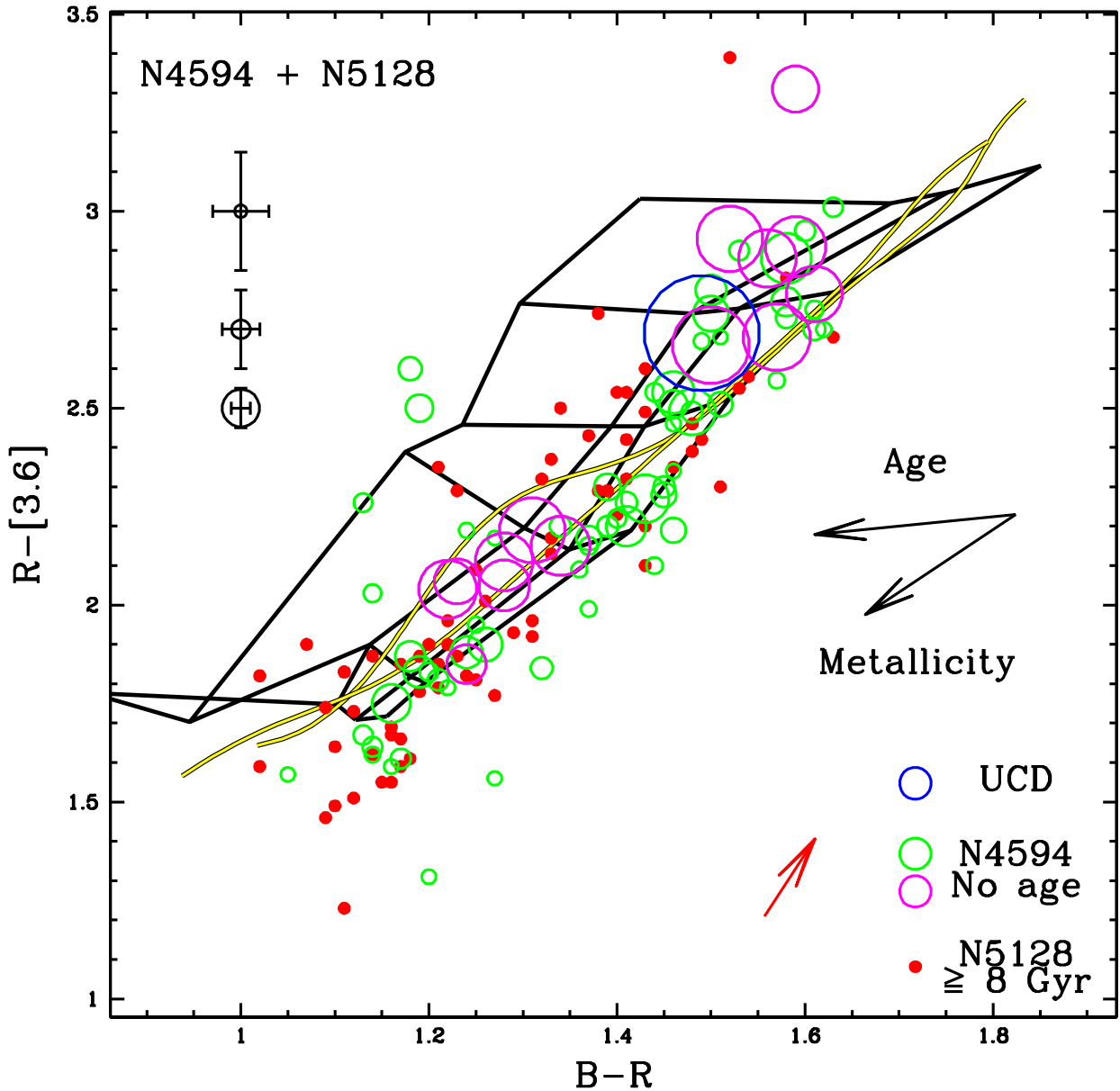}}
\caption{NGC~4594 and NGC~5128 GCs \rf versus B--R colour-colour diagram.  Layout is like Fig.~\ref{figurrch1} and again the circles are inversely proportional to photometric uncertainties.  NGC~4594 GCs are shown as green circles.  Bright (\f $<16.6$ mag.) NGC~4594 GCs are highlighted in magenta and the UCD associated with NGC~4594 is blue.  Although the NGC~4594 GCs occupy the $\sim7$ Gyr region on the SSP grids, analysis in Section~\ref{models} suggests the model predictions at high metallicities are too red in \br, thus no significant age-substructure can be reported.  The Y06 model predictions are presented in yellow for a 13 Gyr (curvy line) and 11 Gyr SSP.  See text for discussion.}\label{figbrrch1}
\end{figure*}

With these caveats in mind, the NGC~4594 GCs are presented in \rf versus B--R colour-colour space in Figure~\ref{figbrrch1}.  Analysis of Figure~\ref{figspitcolourres} in Section~\ref{models} provides no strong preference for a particular SSP to be employed in the \rf versus B--R diagram, thus the CB08 models used in Figure~\ref{figurrch1} are used again.  It was also shown in Section~\ref{models} that, relative to the data, all models predict lower $B-$band fluxes relative to the GC data at high-metallicities, hence GCs will appear younger than they actually are on the \rf versus B--R diagram.  Indeed, in Figure~\ref{figbrrch1} the NGC~5128 GCs with high metallicities tend to occupy younger regions in the SSP model grid, even though they have spectroscopic ages of $\ge8$ Gyr.

The NGC~4594 GCs show a significant (i.e. greater than the photometric uncertainties) spread of colours at \rf $\ga2.5$.  This could result from an intrinsic age substructure among the metal-rich GCs.  Given the possibility that the SSP models under-predict the $B-$band flux at high metallicities, makes any attempt to interpret the apparent colour-spread as an age effect very uncertain.  If the data and models were taken at face value, the CB08 predictions suggest the metal-rich GCs span an age range of $\sim5-12$ Gyrs.  Previous studies have found no evidence for a population of intermediate-aged GCs.  Larsen et al. (2002) analysed absorption line indices of $\sim6$ metal-rich NGC~4594 GCs and found ages of $\ga10$ Gyr.  Hempel et al. (2007) analysed K-band data of $\sim1$ dozen NGC~4594 metal-rich GCs in a similar manner and found no evidence for age substructure.

At \rf $\sim2.1$ in Figure~\ref{figbrrch1} a subpopulation of ``abnormally luminous'' blue GCs (Spitler et al.~2006) are highlighted.  In colour-colour space, these GCs show bluer optical colours than their fainter GC counterparts, possibly suggesting they have younger ages of approximately $5-7$ Gyrs.  These will be discussed further in the Section~\ref{hotstars} below.

Shown as a blue circle in Figure~\ref{figbrrch1} is the UCD candidate associated with NGC~4594 (see \S\ref{match}).  Despite the excellent photometric data on this object, the discrepancy between the data and SSP models makes it difficult to age-date the UCD with just photometry.  However, it can be concluded that the UCD shows colours consistent with NGC~4594 metal-rich GCs, suggesting its metallicity is just a few tenths of a dex below solar metallicity.

\subsubsection{Hot Horizontal Branch Stars}\label{hotstars}

From intermediate- to low-metallicities, the NGC~4594 GCs in Figure~\ref{figbrrch1} show a non-linear structure in colour-colour space that roughly resembles the shape of Y06 SSP 13 Gyr model predictions (given in the same figure), whose inflection-point results from a changing HB morphology (described in Section~\ref{bimodality}).  Interestingly, the NGC~4594 GCs at the point where the Y06 inflection point becomes apparent (\br $\sim 1.3$ and \rf $\sim2.1$) are dominated by the abnormally luminous blue GCs discussed in Section~\ref{ages}.  If these objects were omitted, the resemblance to the Y06 13 Gyr inflection-point disappears completely.

Although evidence contrary to the scenario of Yoon et al. (2006) has been presented (see Sec.~\ref{bimodality}), it does not rule out their model prediction regarding the effects from a changing HB morphology.  Testing the Y06 models with observations is difficult because few GCs fall into the gap between the GC metallicity subpopulations -- exactly where the predicted inflection point is predicted by Yoon et al. (2006).  Furthermore, a strong dependence on age (compare the 13 and 11 Gyr Y06 predictions in Fig.~\ref{figbrrch1}) makes it possible to tweak models to the observations by carefully picking the exact ages of these ``old'' GCs, since old GCs generally have uncertainties of a few Gyrs.  Thus the present observations can provide no constraint on the possible presence of an inflection point as predicted by the Y06 models.

Another possibility is that these GCs host an exotic HB.  Certain Galactic GCs host HBs that show a unusually large number of very hot stars, giving an extended appearance to the HB in individual GC colour-magnitude diagrams.  These extended HBs tend to be found in more massive GCs in the Galaxy (Recio-Blanco et al. 2006; Lee et al. 2007b) and they will naturally emit more light at bluer wavelengths compared to GCs with normal HBs.  While the \rf colours of the NGC~4594 luminous blue GCs should not be effected by such stars, the \br colours might be.  This combination could lead to a situation resembling the data in Figure~\ref{figbrrch1}.

\begin{figure}
\resizebox{1\hsize}{!}{\includegraphics[angle=0]{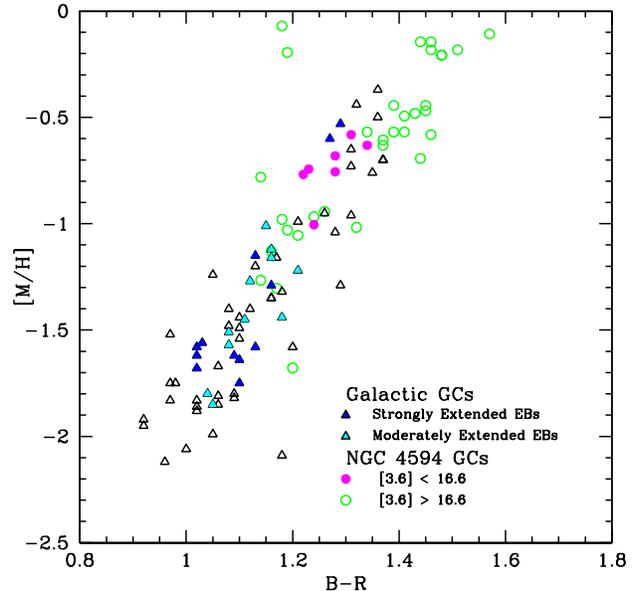}}
\caption{A comparison between Galactic GCs and NGC~4594 GCs.  Galactic data is from Harris (1996) and is restricted to GCs with reddening values E(B--V) $<0.5$.  The NGC~4594 metallicities were estimated using their \rf colours and the [M/H] transformation of Tab.~\ref{tabspitmetal}.  Galactic GCs with strongly extended HB morphologies (from Lee et al. 2007b) are presented with solid, blue triangles.  Moderately extended HBs are highlighted with a cyan triangle.  NGC~4594 GCs brighter and fainter than [3.6] $=16.6$ are presented as solid magenta and open green circles, respectively.  At the B--R colour of the NGC~4594 massive GCs (\br $\sim1.3$), two Galactic GCs (NGC~6388 and NGC~6441; see Rich et al. 1997 and Busso et al. 2007) with strongly extended HBs show systematically bluer colours than other Galactic GCs of a similar [M/H].}\label{figmw}
\end{figure}

To determine whether \br colours are sensitive to extended HB morphologies, Figure~\ref{figmw} highlights Galactic GCs with extended HBs (classified as such by Lee et al.~2007b).  In the Figure, it is apparent that GCs with unusually extended HBs show colours that are roughly consistent with those GCs with normal HBs.  This is in agreement with Smith \& Strader (2007) who found no residual HB trend with optical colours (although they didn't analysis B--R colours) after correlations with metallicity were removed.

In Figure~\ref{figmw} there are two metal-rich Galactic GCs with extended HB morphologies (NGC~6388 and NGC~6441; Rich et al.~1997; Busso et al. 2007) that have \br colours ($\sim1.3$) similar to the abnormally luminous NGC~4594 blue GCs.  The coincidence between these GCs in this parameter space and the systematically lower values of \br for their observed metallicities (as observed in the NGC~4594 GCs) suggests the presence of extended HBs in these NGC~4594 GCs might be remotely plausible.  UV observations are needed for confirmation (e.g. Sohn et al.~2006).

\section{Conclusions}\label{conclusions}

In this work, a catalogue of 260 GC candidates with good optical and Spitzer [3.6] photometry is presented, enabling, for the first time, an examination of the optical to mid-IR properties of a large sample of GCs.  Following previous examples in the literature using K-band photometry, the age-substructure of these GC systems is examined with colour-colour diagrams including the [3.6]--band.  A spread of ages in the NGC~5128 metal-rich GC subpopulation might have been reported from photometry alone.  However, the youngest GCs found in the large spectroscopic study of \citet{2008MNRAS.386.1443B} do not occupy a young region in colour-colour space and actually show photometric properties that are consistent with the old NGC~5128 GCs.  This observation likely demonstrates that the age-metallicity degeneracy is indeed difficult to overcome when using only typical photometric observations.  Spectroscopic age confirmation is required for any age-substructure to be interpreted from old stellar population in such colour-colour diagrams.

For the NGC~4594 GC system, for which no large sample of GCs with spectroscopic ages exists, the metal-rich GCs generally fall onto the the 5--7 Gyr SSP model predictions in \br versus \rf.  However, an empirical comparison between data and SSP models (see below) suggests the models likely under-predict the $B-$band flux relative to the data at these metallicities.  Until this discrepancy is understood, the apparent subpopulation of intermediate-aged NGC~4594 metal-rich GCs will remain unconfirmed.

Another notable feature in the NGC~4594 GC colour-colour diagram (Fig.~\ref{figbrrch1}) is an apparent non-linearity traversal of the parameter space, which resemble the SSP model predictions of S. Yoon (2006, priv. comm.; Y06).  Driving this resemblance are a subpopulation of very luminous and extended GC candidates first noted by Spitler et al. (2006).  These objects have \br colours similar to two Galactic GCs that host very extended HB morphologies (Rich et al. 1997; Busso et al. 2007).  These Galactic and the luminous NGC~4594 GCs are bluer in \br than expected for their metallicities, perhaps suggesting they share similar HB morphologies (see Fig.~\ref{figmw}).

Another possibility is that the massive NGC~4594 GCs are $\sim5$ Gyr younger than the bulk of the GC system.  This would explain their high luminosities, but not the extended sizes they tend to show (Spitler et al. 2006).  A dwarf nuclei with a young stellar component that accreted onto NGC~4594 might match these characteristics, although testing this scenario with simulations is required.

Using the excellent mass proxy, \f, and the largest colour dynamical range studied thus far for subpopulation colour-magnitude trends, the present work demonstrates the NGC~5128 blue GC subpopulation shows redder colours at higher luminosities.  The NGC~4594 sample does not span a large enough range of the GC luminosity function to constrain such trends.  The NGC~5128 blue GCs show a mass-metallicity proportionality of $Z\propto M^{0.19}$; noticeably weaker compared to other galaxies.  This confirms initial suspicions of Spitler et al. (2006) that the strength of the blue tilt decreases with the host galaxy luminosity.  No such trend is found among the red GCs in NGC~5128.  The \citet{2008MNRAS.386.1443B} NGC~5128 GC spectroscopic ages allow, for the first time, the blue tilt to be studied independent of any young to intermediate-aged GC.  It is concluded that age substructure in not likely playing a role in the blue tilt.

The scenario of Yoon et al. (2006) is tested with the \rf GC data.  Such colours are insensitive to hot HB stars (see the Y06 models in Fig~\ref{figspitmetal}), hence can determine whether GC colour bimodality implies metallicity bimodality without effects from a rapidly changing HB morphology.  While the NGC~4594 sample is too small to constrain this scenario, the old GCs in NGC~5128 provide strong evidence for \rf colour bimodality, which can safely be interpreted as metallicity bimodality.  This confirms the spectroscopic analysis of NGC~5128 GCs by \citet{2008MNRAS.386.1443B}.  Strader et al. (2007) and Kundu \& Zepf (2007) provide further evidence for GC system metallicity bimodality in massive ellipticals, suggesting the Yoon et al. (2006) scenario does not apply to extragalactic (nor Galactic) GC systems.

To conclude, despite a low IRAC pixel resolution and relatively short exposure times (e.g. 240s -- NGC~4594, 72s -- NGC~5128), the present work demonstrates that interesting science in the field of extragalactic GC system astronomy is indeed possible with Spitzer IRAC imaging.  The largest prerequisite for such work is a good optical GC catalogue, because Spitzer photometry alone cannot be used for contamination removal for such faint objects.  

\section*{Acknowledgments}

The authors appreciate useful discussions with R. Proctor and the enthusiasm and assistance provided by the stellar population modellers:  M. Cantiello, S. Charlot, C. Maraston, and S. Yoon.  The referee provided a number of very useful comments that helped improve the discussion.  DF thanks the Australian Research Council for financial support.  This work is based on observations made with the Spitzer Space Telescope, which is operated by the Jet Propulsion Laboratory, California Institute of Technology under a contract with NASA.

\label{lastpage}

\end{document}